\documentclass[10pt]{sigplanconf-pldi16}
\usepackage{SIunits}            % typset units correctly
\usepackage{courier}            % standard fixed width font
\usepackage[scaled]{helvet} % see www.ctan.org/get/macros/latex/required/psnfss/psnfss2e.pdf
\usepackage{url}                  % format URLs
\usepackage{listings}          % format code
\usepackage{enumitem}      % adjust spacing in enums
\usepackage[colorlinks=true,allcolors=blue,breaklinks,draft=false]{hyperref}   % hyperlinks, including DOIs and URLs in bibliography
\newcommand{\doi}[1]{\href{http://dx.doi.org/#1}{doi:~\Hurl{#1}}}   % print a hyperlinked DOI
\newcommand{\arxiv}[1]{\href{http://arxiv.org/abs/#1}{arXiv:~\Hurl{#1}}}   % print a hyperlinked DOI

\usepackage{amsmath,amsthm}
\usepackage{graphicx}
\usepackage{tikz}
\usepackage{color}
\usepackage{stmaryrd}
\usepackage{eufrak}
\usepackage{scalerel}
\usepackage{stackengine,wasysym}

\theoremstyle{theorem}
\newtheorem{theorem}{Theorem}[section]
\theoremstyle{definition}

\begin{document}

\title{Parameterized Local Lexing}

%
% any author declaration will be ignored  when using 'pldi' option (for double blind review)
%

%\authorinfo{.}
\authorinfo{Steven Obua}
{Recursive Mind (www.recursivemind.com)}
{steven@recursivemind.com}

\maketitle

\newcommand{\parameters}{\ensuremath{\Phi}}
\newcommand{\nonterminals}{\ensuremath{\mathfrak{N}}}
\newcommand{\terminals}{\ensuremath{\mathfrak{T}}}
\newcommand{\startsymbol}{\ensuremath{\mathfrak{S}}}
\newcommand{\startparameter}{\ensuremath{\phi_{\textsl{start}}}}
\newcommand{\rules}{\ensuremath{\mathfrak{R}}}
\newcommand{\lexer}{\ensuremath{\textsl{Lex}}}
\newcommand{\selector}{\ensuremath{\textsl{Sel}}}
\newcommand{\lang}{\ensuremath{\mathcal{L}}}
\newcommand{\prefixlang}{\ensuremath{\lang_\text{prefix}}}
\newcommand{\derives}{\ensuremath{\overset{*\ }{\Rightarrow}}}
\newcommand{\lderives}{\ensuremath{\ {\overset{*\ }{\Rightarrow}}_L}\ }
\newcommand{\terminalsof}[1]{\ensuremath{[#1]}}
\newcommand{\charsof}[1]{\ensuremath{\overline{#1}}}
\newcommand{\locallexing}{\ensuremath{\ell\ell}}
\newcommand{\lextoken}[2]{\small\ensuremath{\dfrac{\texttt{#1}}{\textsl{#2}}}}
\newcommand{\sele}{\ensuremath{<_\selector}}
\newcommand{\nterminal}[1]{\ensuremath{\textsl{#1}}}
\newcommand{\terminal}[1]{\ensuremath{\textsl{#1}}}
\newcommand{\gsymbol}[1]{\ensuremath{\textsl{#1}}}
\newcommand{\closure}[3]{\ensuremath{\textsl{closure}_{#1}\, #2\ #3}}
\newcommand{\emptyseq}{\ensuremath{\varepsilon}}
\newcommand{\seqlen}[1]{\ensuremath{\left|{#1}\right|}}
\newcommand{\charslen}[1]{\seqlen{\charsof{#1}}}
\newcommand{\paths}[2]{\ensuremath{\mathcal{P}_{#1}^{#2}}}
\newcommand{\tokensat}[1]{\ensuremath{\mathcal{X}_{#1}}}
\newcommand{\admissibletokens}[2]{\ensuremath{\mathcal{W}_{#1}^{#2}}}
\newcommand{\selectedtokens}[2]{\ensuremath{\mathcal{Z}_{#1}^{#2}}}
\newcommand{\appendtokens}[1]{\ensuremath{\operatorname{Append}_{#1}}}
\newcommand{\limit}[2]{\ensuremath{\operatorname{limit}\ {#1}\ {#2}}}
\newcommand{\allpaths}{\ensuremath{\mathfrak{P}}}
\newcommand{\allitems}{\ensuremath{\mathfrak{I}}}
\newcommand\pto{\mathrel{\ooalign{\hfil$\mapstochar\mkern5mu$\hfil\cr$\to$\cr}}}
\newcommand{\induced}[1]{\ensuremath{\overline{#1}}}
\newcommand{\outputs}[1]{\ensuremath{\sigma({#1})}}
\newcommand{\itemdot}{\ensuremath{\makebox[7pt]{\textbullet}}}
\newcommand{\someitems}{\ensuremath{I}}

\begin{abstract}
Building on the concept of \emph{local lexing} the concept of \emph{parameterized local lexing} is introduced.
\end{abstract}
\section{Motivation}
We previously introduced \emph{local lexing} as a new semantics for integrated lexing and parsing, and provided an algorithm (based on Earley's Algorithm) for it~\cite{locallexing}. This algorithm can handle arbitrary context-free grammars. In principle it can even deal with infinite grammars, but to harness this capability in practice we need a finite representation of such possibly infinite grammars. \emph{Parameterized local lexing} delivers such a finite representation, and comes with an algorithm tuned to that representation. Of course the finite representation as presented in this paper is not the only possible one, but it is simple and intuitive and yet powerful.

\section{Definition}
We call a tuple \[(\parameters, \nonterminals, \terminals, \rules, \startsymbol, \startparameter, \Sigma, \lexer, \selector)\] a parameterized local lexing iff:
\begin{itemize}
\item $\parameters$ is a non-empty set of parameters.
\item $\nonterminals$ and $\terminals$ are disjoint sets of nonterminals and terminals.
\item $\startsymbol \in \nonterminals$ and $\startparameter \in \parameters$ are the start symbol and start parameter, respectively.
\item $\rules$ is a set of parameterized rules. Each rule has the form 
  \[ \nterminal{N}_{f_{k+1}} \rightarrow \gsymbol{X}_1^{f_1} \ldots \gsymbol{X}_k^{f_k}  \]
  where $\nterminal{N}, \gsymbol{X}_1, \ldots, \gsymbol{X}_k \in \nonterminals$, and 
  $f_1, \ldots, f_k, f_{k+1}$
  are partial functions with the signatures $f_i : \parameters^{2i-1} \pto \parameters$ for $i \in \{1, \ldots, k+1\}$.
\item $\Sigma$ is a set of characters.
\item $\lexer$ is a function which for each $(t, \alpha, D, k)$ such that
$t \in \terminals$, $\alpha \in \parameters$, $D \in \Sigma^*$ and $k \in \{0, \ldots, |D|\}$ returns a set consisting of tuples 
$(t, \alpha, \beta, c)$ such that $\beta \in \parameters$, $c \in \Sigma^*$, $k + |c| \leq |D|$ and
$c_i = D_{k + i}$ for all $i$ such that $0 \le i \le |c| - 1$.
\item $\selector$ takes two sets $A, B \subseteq \terminals \times \parameters \times \parameters \times \Sigma^*$ 
such that $A \subseteq B$ and returns a set 
$\selector(A, B)$ such that $A \subseteq \selector(A, B) \subseteq B$. 
\end{itemize}

\newcommand{\failsymbol}{\bot}
\newcommand{\startnonterminal}{\top}

\section{Semantics}

\newcommand{\paramhull}[1]{\ensuremath{\langle{#1}\rangle}}
\newcommand{\itemparams}{\rho}
\newcommand{\altitemparams}{\xi}
\newcommand{\someitem}{\ensuremath{x}}
\newcommand{\itemsi}[1]{\ensuremath{\mathcal{I}_{#1}}}
\newcommand{\itemsj}[2]{\ensuremath{\mathcal{J}_{#1}^{#2}}}
\newcommand{\ctokens}[2]{\ensuremath{\mathcal{T}_{#1}^{\,#2}}}
\newcommand{\tokenbij}{\mathcal{U}}

The semantics of parameterized local lexing is provided by translating it to ordinary local lexing.
The result is a context-free grammar $(\induced{\nonterminals}, \induced{\terminals}, \induced{\rules}, \startnonterminal)$
together with a lexer $\induced{\lexer}$ and selector $\induced{\selector}$ operating on the same character set $\Sigma$ as the parameterized local lexing. The resulting nonterminals and terminals of the new grammar are defined as follows: 
\[
\begin{array}{ccl}
  \induced{\nonterminals} & = & (\nonterminals \times \parameters \times \parameters) \cup \{\startnonterminal, \failsymbol\}\\
  \induced{\terminals} & = & \terminals \times \parameters \times \parameters 
\end{array}
\] 
For elements $(N, \alpha, \beta) \in \induced{\nonterminals}$ and $(t, \alpha, \beta) \in \induced{\terminals}$ we also 
write $N_\beta^\alpha$ and $t_\beta^\alpha$, respectively. Here $\alpha$ is called the \emph{input parameter}, and $\beta$ is called
the \emph{output parameter}.
The nonterminal $\failsymbol$ represents failure and therefore does not appear on the left hand side of any rule in \induced{\rules}. 
The nonterminal $\startnonterminal$ is the start symbol of the new grammar and appears only on the left hand side of the rules in the following subset of $\induced{\rules}$:
\[ \left\{ \startnonterminal \rightarrow \startsymbol^{\startparameter}_{\beta} \mid \beta \in \parameters \right\} \]

Given $u \geq 1$ partial functions $f_1, \ldots, f_u$ with signatures $f_i : \parameters^{2i-1} \pto \parameters$ for $i \in \{1, \ldots, u\}$, we define the set
$\paramhull{f_1, \ldots, f_u}$ to consist of all sequences $\itemparams \in \parameters^{2 u}$ such that 
\[
  \begin{array}{l}
   \phantom{\wedge \ } \text{$f_1$ is defined at $\itemparams_0$} \\
   \wedge \ \itemparams_1 = f_1(\itemparams_0) \\
   \wedge \ \text{$f_2$ is defined at $(\itemparams_0, \itemparams_1, \itemparams_2)$} \\
   \wedge \ \itemparams_3 = f_2(\itemparams_0, \itemparams_1, \itemparams_2) \\
   \wedge \ \text{$f_3$ is defined at $(\itemparams_0, \itemparams_1, \itemparams_2, \itemparams_3, \itemparams_4)$} \\
   \wedge \ \itemparams_5 = f_3(\itemparams_0, \itemparams_1, \itemparams_2, \itemparams_3, \itemparams_4) \\
   \vdots \\
   \wedge \ \text{$f_{u}$ is defined at $(\itemparams_0, \itemparams_1, \ldots, \itemparams_{2u-2})$} \\
   \wedge \ \itemparams_{2u-1} = f_{u}(\itemparams_0, \itemparams_1, \ldots, \itemparams_{2u-2})
  \end{array}
\]

With each parameterized rule $r = \nterminal{N}_{f_{k+1}} \rightarrow \gsymbol{X}_1^{f_1} \ldots \gsymbol{X}_k^{f_k}$ we associate
its induced set of rules $\induced{r} \subseteq \induced{\rules}$. Firstly, $\induced{r}$ contains all rules
\[  \nterminal{N}_\beta^{\,\alpha} \rightarrow 
  {\gsymbol{X}_1}_{\beta_1}^{\alpha_1} \ldots {\gsymbol{X}_k}_{\beta_k}^{\alpha_k}\]
such that $\alpha\, \alpha_1\,\beta_1 \ldots \alpha_k\, \beta_k\, \beta \in \paramhull{f_1, \ldots, f_{k+1}}$.

Secondly, we also need to somehow take into account those cases where some $f_i$ happens to be undefined on its arguments. 
While these cases do not affect the language $\lang$
of the induced context-free grammar, they possibly do affect its prefix language $\prefixlang$. We choose to let $\induced{r}$ contain also all rules
\[  \nterminal{N}_\beta^{\,\alpha} \rightarrow 
  {\gsymbol{X}_1}_{\beta_1}^{\alpha_1} \ldots {\gsymbol{X}_{h}}_{\beta_{h}}^{\alpha_{h}}\ \failsymbol\]
such that
\[
  \begin{array}{l}
   h \in \{1, \ldots, k\} \wedge \beta \in \parameters \wedge \beta_h \in \parameters \\
   \wedge\ \alpha\, \alpha_1\, \beta_1 \ldots \alpha_h \in \paramhull{f_1, \ldots, f_h} \\
   \wedge \ \text{$f_{h+1}$ is undefined at $(\alpha, \alpha_1, \beta_1, \ldots, \alpha_{h}, \beta_{h})$}
  \end{array}
\]
The induced set of rules $\induced{\rules}$ is thus defined as
\[ \induced{\rules} = \left\{ \startnonterminal \rightarrow \startsymbol^{\startparameter}_{\beta} \mid \beta \in \parameters \right\} \cup \bigcup\limits_{r \in \rules} \induced{r}\]

Defining the induced lexer is straightforward:
\[ \induced{\lexer}(t^\alpha_\beta)(D, k) = \left\{ (t^\alpha_\beta, c) \mid (t, \alpha, \beta, c) \in \lexer(t, \alpha, D, k) \right\}  \]
There is an obvious bijection $\tokenbij : \terminals \times \parameters \times \parameters \times \Sigma^* \rightarrow \induced{\terminals} \times \Sigma^*$
between parameterized tokens and ordinary tokens, given by $\tokenbij (t, \alpha, \beta, c) = (t^\alpha_\beta, c)$. Lifted to a bijection between token sets, we use it to define the induced selector:
\[\induced{\selector}(A, B) = \tokenbij (\selector(\tokenbij^{-1}(A), \tokenbij^{-1}(B))) \]

A character sequence $D \in \Sigma^*$ is said to be in the character language of the parameterized local lexing iff $\locallexing(D) \neq \emptyset$ with respect to the induced local lexing. Furthermore, we can associate with each $D \in \Sigma^*$ the set of its outputs $\outputs{D} \subseteq \parameters$ via
\[ \outputs{D} = \left\{ \beta \mid \exists\ p.\ p \in \locallexing(D) \text{ and } \startnonterminal \Rightarrow \startsymbol^{\startparameter}_{\beta} \derives \terminalsof{p}   \right\}. \]
Obviously, $\locallexing(D) = \emptyset$ iff $\outputs{D} = \emptyset$.

\section{Algorithm}
Given $\parameters$, $\nonterminals$, $\terminals$ and $\rules$ are all finite, parameterized local lexing can be implemented simply by reducing it via its semantics to ordinary local lexing, and then applying the Earley-based local lexing algorithm (ELLA). This is rather awkward though as
the induced grammar is potentially much larger than the original grammar. In particular, $|\induced{\rules}|$ might be much larger than $|\rules|$. Furthermore, we would also like to have the option of doing parameterized local lexing for an infinite set of parameters $\parameters$ like the set of natural numbers, but usually this would lead to an infinite set of rules $\induced{\rules}$.

Luckily it is straightforward to derive from ELLA an algorithm for parameterized local lexing (PELLA) which avoids such a translation to ordinary local lexing but works directly on \emph{parameterized items}. A parameterized item is a tuple $(r, d, i, j, \itemparams)$, 
such that $r = \nterminal{N}_{f_{k+1}} \rightarrow \gsymbol{X}_1^{f_1} \ldots \gsymbol{X}_k^{f_k} \in \rules$, $0 \leq d \leq k$ and $0 \leq i \leq j \leq |D|$, where $D \in \Sigma^*$ is the input under consideration. The sequence $\itemparams \in \parameters^{2(d+1)}$ records the choices of parameters that have been made so far, and therefore we demand that
the invariant $\itemparams \in \paramhull{f_1, \ldots, f_{d+1}}$ holds for each parameterized item.

Before describing how PELLA operates on parameterized items, let us establish a correspondence between parameterized items and the items of the induced ordinary local lexing. Just how each parameterized rule corresponds to a set of ordinary rules, each parameterized item $\someitem$ corresponds to a set of ELLA items $\induced{\someitem}$. Firstly, for each rule 
$q = \nterminal{N}_\beta^{\,\alpha} \rightarrow {\gsymbol{X}_1}_{\beta_1}^{\alpha_1} \ldots {\gsymbol{X}_k}_{\beta_k}^{\alpha_k} \in \induced{r}$ the set $\induced{\someitem}$ contains all items
$(q, d, i, j)$ such that  
\[\operatorname{take}_{2(d+1)}\ (\alpha, \alpha_1, \beta_1, \alpha_2, \beta_2, \ldots, \alpha_{k}, \beta_{k}, \beta) = \itemparams. \]
Here $\operatorname{take}_n s$ denotes the sequence resulting from taking the first $n$ elements of the sequence $s$. Secondly, for each rule 
$q = \nterminal{N}_\beta^{\,\alpha} \rightarrow 
  {\gsymbol{X}_1}_{\beta_1}^{\alpha_1} \ldots {\gsymbol{X}_{h}}_{\beta_{h}}^{\alpha_{h}}\ \failsymbol \in \induced{r}$ the set $\induced{\someitem}$ contains all items
$(q, d, i, j)$ such that $d \leq h-1$ and
\[\operatorname{take}_{2(d+1)}\ (\alpha, \alpha_1, \beta_1, \alpha_2, \beta_2, \ldots, \alpha_{h}) = \itemparams. \]
\begin{theorem}\label{th:nonemptyinduceditems}
Let $x$ be a parameterized item. Then $\induced{x} \neq \emptyset$.
\end{theorem}
\begin{proof}
Let $x = (r, d, i, j, \itemparams)$, and $r = \nterminal{N}_{f_{k+1}} \rightarrow \gsymbol{X}_1^{f_1} \ldots \gsymbol{X}_k^{f_k}$. 
We know that $\itemparams \in \paramhull{f_1, \ldots, f_{d+1}}$.

If there is $\itemparams' \in \parameters^{2(k - d)}$ such that
$\altitemparams = \itemparams \itemparams' \in \paramhull{f_1, \ldots, f_{k+1}}$ then $\induced{x}$ is not empty as
$(\nterminal{N}^{\altitemparams_0}_{\altitemparams_{2k+1}} \rightarrow 
\gsymbol{X}^{\altitemparams_1}_{\altitemparams_2} \ldots \gsymbol{X}^{\altitemparams_{2k-1}}_{\altitemparams_{2k}}, d, i, j) \in \induced{x}$. 
Otherwise there must be $h$ with $d + 1 \leq h < k$ and $\itemparams' \in \parameters^{2(h-d-1)}$ such that 
$\altitemparams = \itemparams \itemparams' \in \paramhull{f_1, \ldots, f_h}$, together with some $\gamma \in \parameters$ such that
$f_{h+1}$ is undefined at $\altitemparams\,\gamma$. This means that for any $\beta \in \parameters$, we have 
$(\nterminal{N}^{\altitemparams_0}_{\beta} \rightarrow 
\gsymbol{X}^{\altitemparams_1}_{\altitemparams_2} \ldots \gsymbol{X}^{\altitemparams_{2h-1}}_{\gamma} \failsymbol, d, i, j) \in \induced{x}$. 
\end{proof}

Given an item set $\someitems$, we define its induced item set $\induced{\someitems}$ as
\[ \induced{\someitems} = \bigcup\limits_{w \in \someitems} \induced{w} \]
\begin{figure}
\begin{align*}
&\operatorname{Init}  = \\
&\{  (\startsymbol_{f_{u+1}} \rightarrow \itemdot \gsymbol{X}_1^{f_1} \ldots \gsymbol{X}_u^{f_u}, 0, 0, 
\startparameter\  f_1(\startparameter)) \mid \\
&\phantom{\wedge}\ (\startsymbol_{f_{u+1}} \rightarrow \gsymbol{X}_1^{f_1} \ldots \gsymbol{X}_u^{f_u}) \in \rules \\
& \wedge\ \text{$f_1$ is defined at $\startparameter$} \} \\ \\
&\operatorname{Predict}\  k\ \someitems = \someitems\ \cup\\
 &\{  (\nterminal{M}_{g_{u+1}} \rightarrow \itemdot \gsymbol{Y}_1^{g_1} \ldots \gsymbol{Y}_u^{g_u}, k, k, 
 \itemparams_{2d+1}\  g_1(\itemparams_{2d+1})) \mid \\
& \phantom{\wedge}\ (\nterminal{M}_{g_{u+1}} \rightarrow \gsymbol{Y}_1^{g_1} \ldots \gsymbol{Y}_u^{g_u}) \in \rules\\
& \wedge\ \text{$g_1$ is defined at $\itemparams_{2d+1}$} \\ 
& \wedge (N_{f_{v+1}} \rightarrow \gsymbol{X}_1^{f_1} \ldots \gsymbol{X}_d^{f_d} \itemdot \gsymbol{X}_{d+1}^{f_{d+1}} \ldots \gsymbol{X}_v^{f_v}, 
i, k, \itemparams) \in \someitems \\
& \wedge \gsymbol{X}_{d+1} = \nterminal{M} \}\\ \\
&\operatorname{Complete}\   k\ \someitems = \someitems\ \cup\\
 &\{  (N_{f_{v+1}} \rightarrow \gsymbol{X}_1^{f_1} \ldots \gsymbol{X}_d^{f_d} \gsymbol{X}_{d+1}^{f_{d+1}}\itemdot 
 \ldots \gsymbol{X}_v^{f_v}, i, k, 
 \itemparams'\, f_{d+2}(\itemparams')) \mid \\
    &\phantom{\wedge}\ (N_{f_{v+1}} \rightarrow \gsymbol{X}_1^{f_1} \ldots \gsymbol{X}_d^{f_d} \itemdot \gsymbol{X}_{d+1}^{f_{d+1}} \ldots \gsymbol{X}_v^{f_v}, i, j, \itemparams) \in \someitems \\ 
&  \wedge (\nterminal{M}_{g_{u+1}} \rightarrow \gsymbol{Y}_1^{g_1} \ldots \gsymbol{Y}_u^{g_u} \itemdot, j, k, \altitemparams) 
\in \someitems \\ 
& \wedge \gsymbol{X}_{d+1} = M \wedge \itemparams_{2d+1} = \altitemparams_{0}
\wedge \itemparams' = \itemparams\, \altitemparams_{2u+1} \\
&\wedge\ \text{$f_{d+2}$ is defined at $\itemparams'$} \} \\ \\ 
&\operatorname{Tokens}\   T\ k\ \someitems = \selector\ T\  \\
&\{ x \mid (N_{f_{v+1}} \rightarrow \gsymbol{X}_1^{f_1} \ldots \gsymbol{X}_d^{f_d} \itemdot \gsymbol{X}_{d+1}^{f_{d+1}} \ldots \gsymbol{X}_v^{f_v},
i, k, \itemparams) \in \someitems \\
     & \wedge \gsymbol{X}_{d+1} \in \terminals \wedge x \in \lexer(\gsymbol{X}_{d+1}, \itemparams_{2d+1}, D, k)\}\\ \\
&\operatorname{Scan}\   T\ k\ \someitems = \someitems\ \cup\\
& \{  (\nterminal{N}_{f_{v+1}} \rightarrow \gsymbol{X}_1^{f_1} \ldots \gsymbol{X}_d^{f_d} 
      \gsymbol{X}_{d+1}^{f_{d+1}} \itemdot \ldots \gsymbol{X}_v^{f_v}, i, k + |c|,
      \itemparams'\, f_{d+2}(\itemparams')) \mid \\
    & \phantom{\wedge}\ (\nterminal{N}_{f_{v+1}} \rightarrow \gsymbol{X}_1^{f_1} \ldots \gsymbol{X}_d^{f_d} \itemdot 
      \gsymbol{X}_{d+1}^{f_{d+1}} \ldots \gsymbol{X}_v^{f_v}, i, k, \itemparams) \in I\\
    & \wedge\ (\gsymbol{X}_{d+1}, \itemparams_{2d+1}, \beta, c) \in T \wedge \itemparams' = \itemparams\, \beta\\
    & \wedge\ \text{$f_{d+2}$ is defined at $\itemparams'$} \} 
\end{align*}
\caption{Building Blocks of PELLA}
\label{fig:pellaoperators}
\end{figure}

\newcommand{\norm}[1]{\ensuremath{\operatorname{Norm}\left(#1\right)}}

The idea of PELLA is that it operates on item sets $\someitems$ just as ELLA would operate on $\induced{\someitems}$. For each of the original building blocks 
$\operatorname{Init}'$, $\operatorname{Predict}'$, $\operatorname{Complete}'$, $\operatorname{Tokens}'$ and $\operatorname{Scan}'$ of ELLA we define the corresponding 
building blocks $\operatorname{Init}$, $\operatorname{Predict}$, $\operatorname{Complete}$, $\operatorname{Tokens}$ and $\operatorname{Scan}$ of PELLA as shown in Figure~\ref{fig:pellaoperators}
 (the use of the $\itemdot$ notation for PELLA items is analogous to its use for ELLA items).
Apart from this adaptation of the building blocks to a parameterized setting, PELLA is defined via exactly the same equations as ELLA is, as shown in Figure~\ref{fig:pella}.
\begin{figure}
\begin{align*}
\limit f X &= \bigcup\limits_{n = 0}^\infty f^n(X)\\
\pi_k\,T\,\someitems & = \limit {(\operatorname{Scan} T\ k \circ \operatorname{Complete} k \circ \operatorname{Predict} k)}{\someitems}\\
\itemsj 0 0 & = \pi_0\ \emptyset\ \operatorname{Init}\\
\itemsj k {u+1} & = \pi_k\ \ctokens k {u+1}\ \itemsj k u\\
\itemsi k & = \bigcup\limits_{u=0}^\infty \itemsj k u\\
\itemsj {k+1} 0 & = \pi_{k+1}\ \emptyset{}\ \itemsi k\\
\ctokens k 0 & = \emptyset\\
\ctokens {k} {u+1} & = \operatorname{Tokens}\ \ctokens k u \  k\ \itemsj k u\\
\allitems & = \itemsi {|D|}
\end{align*}
\caption{PELLA equations}
\label{fig:pella}
\end{figure}
\begin{theorem}[Correctness of PELLA]\label{th:correctness:PELLA}
\begin{align*}
\outputs{D} = \{ \itemparams_{2 u + 1} \mid & (\startsymbol_{f_{u+1}} \rightarrow \gsymbol{X}_1^{f_1} \ldots \gsymbol{X}_u^{f_u} \itemdot, 0, |D|, \itemparams) \in \allitems \\
 & \wedge \itemparams_0 = \startparameter \}
\end{align*}
\end{theorem}
Indeed, we can prove the correctness of PELLA by exploiting the fact that PELLA simulates ELLA. The proof works by explaining in detail what we mean by this simulation.
To this end, we need a notion of correspondence between PELLA item sets $\someitems$ and ELLA item sets $\someitems'$. Intuitively, $\induced{\someitems} = \someitems'$ is an obvious candidate for 
such a relation, but this does not work: $\someitems'$ may contain items of the form $(\startnonterminal \rightarrow w_1 \itemdot w_2, i, j)$ or
$(N \rightarrow w \itemdot \failsymbol, i, j)$, both of which will never be in $\induced{\someitems}$. Therefore, we define the correspondence $\someitems \sim \someitems'$ instead as $\induced{\someitems} = \norm{\someitems'}$
where $\norm{\someitems'} = \left\{ (N \rightarrow w_1 \itemdot w_2, i, j) \in \someitems' \mid N \neq \startnonterminal \wedge w_2 \neq \failsymbol \right\}$. 
Indeed, with this choice of correspondence the formulas in Figure~\ref{fig:simulation} hold and show that the building blocks of PELLA simulate the building blocks of ELLA. 
\begin{figure}
\begin{align*}
\operatorname{Init} &\sim \operatorname{Predict}' 0\,\operatorname{Init}' \\
\someitems \sim \someitems' \Rightarrow \operatorname{Predict} k\,\someitems &\sim \operatorname{Predict}' k\,\someitems'\\
\someitems \sim \someitems' \Rightarrow \operatorname{Complete} k\,\someitems &\sim \operatorname{Complete}' k\,\someitems'\\
\someitems \sim \someitems' \Rightarrow \tokenbij(\operatorname{Tokens} T\,k\,\someitems) &= \operatorname{Tokens}'\, \tokenbij(T)\,k\,\someitems' \\
\someitems \sim \someitems' \Rightarrow \operatorname{Scan} T\, k\,\someitems &\sim \operatorname{Scan}'\, \tokenbij(T)\, k\,\someitems'
\end{align*}
\caption{PELLA simulates ELLA}
\label{fig:simulation}
\end{figure}
In the following we will prove the formulas in Figure~\ref{fig:simulation}.

\begin{theorem}\label{th:sim:Init}
\[\operatorname{Init} \sim \operatorname{Predict}' 0\,\operatorname{Init}'\]
\end{theorem}
\begin{proof}
From the definitions it follows that
\begin{align*}
&\norm{\operatorname{Predict}' 0\,\operatorname{Init}'} = \\
&\quad\left\{ (\startsymbol^{\startparameter}_\beta \rightarrow \itemdot w, 0, 0) \mid 
(\startsymbol^{\startparameter}_\beta \rightarrow w) \in \induced{\rules}\right\}
\end{align*}
We first show $\norm{\operatorname{Predict}' 0\,\operatorname{Init}'} \subseteq \induced{\operatorname{Init}}$. 
So assume that $(\startsymbol^{\startparameter}_\beta \rightarrow \itemdot w, 0, 0) \in \norm{\operatorname{Predict}' 0\,\operatorname{Init}'}$. There must be $r \in \rules$ such that
$(\startsymbol^{\startparameter}_\beta \rightarrow w) \in \induced{r}$, and $r = \startsymbol_{f_{k+1}} \rightarrow \gsymbol{X}_1^{f_1} \ldots \gsymbol{X}_k^{f_k}$ for some $f_1, \ldots, f_{k+1}$
and some $\gsymbol{X}_1, \ldots, \gsymbol{X}_k$. 

If $w$ does not end with $\failsymbol$, we know there must be $\alpha_1, \ldots, \alpha_k$ and $\beta_1, \ldots, \beta_k$ such that
$(\startsymbol^{\startparameter}_\beta \rightarrow w) = (\startsymbol_\beta^{\,\startparameter} \rightarrow 
{\gsymbol{X}_1}_{\beta_1}^{\alpha_1} \ldots {\gsymbol{X}_k}_{\beta_k}^{\alpha_k})$, and we also know then that $f_1$ is defined on $\startparameter$, and that 
either $k = 0 \wedge \beta = f_1(\startparameter)$ or  $k > 0 \wedge \alpha_1 = f_1(\startparameter)$. 
For the parameterized item $x = (\startsymbol_{f_{k+1}} \rightarrow \itemdot \gsymbol{X}_1^{f_1} \ldots \gsymbol{X}_k^{f_k}, 0, 0, \startparameter\, f_1(\startparameter))$ this implies
that $x \in \operatorname{Init}$, and that $(\startsymbol^{\startparameter}_\beta \rightarrow \itemdot w, 0, 0) \in \induced{x} \subseteq \induced{\operatorname{Init}}$. 

If on the other hand $w$ ends with $\failsymbol$, then $k > 0$ and there must be $h \in \{1, \ldots, k\}$, $\alpha_1, \ldots, \alpha_h$ and $\beta_1, \ldots, \beta_h$ such that
$(\startsymbol^{\startparameter}_\beta \rightarrow w) = (\startsymbol_\beta^{\,\startparameter} \rightarrow 
{\gsymbol{X}_1}_{\beta_1}^{\alpha_1} \ldots {\gsymbol{X}_h}_{\beta_h}^{\alpha_h} \failsymbol)$. We also know then that $f_1$ is defined on $\startparameter$, and that 
$\alpha_1 = f_1(\startparameter)$. For the parameterized item $x = (\startsymbol_{f_{k+1}} \rightarrow \itemdot \gsymbol{X}_1^{f_1} \ldots \gsymbol{X}_k^{f_k}, 0, 0, \startparameter\, f_1(\startparameter))$ this implies that $x \in \operatorname{Init}$, and that $(\startsymbol^{\startparameter}_\beta \rightarrow \itemdot w, 0, 0) \in \induced{x} \subseteq \induced{\operatorname{Init}}$.

We now prove $\induced{\operatorname{Init}} \subseteq \norm{\operatorname{Predict}' 0\,\operatorname{Init}'}$. So assume $x' \in \induced{\operatorname{Init}}$. 
Then there must be $x \in \operatorname{Init}$ such that $x' \in \induced{x}$. We know that \[x = (\startsymbol_{f_{u+1}} \rightarrow \itemdot \gsymbol{X}_1^{f_1} \ldots \gsymbol{X}_u^{f_u}, 0, 0, 
\startparameter\  f_1(\startparameter))\] for some $\gsymbol{X}_1, \ldots, \gsymbol{X}_u$ and some $f_1, \ldots, f_u$. From this $x' \in \norm{\operatorname{Predict}' 0\,\operatorname{Init}'}$
immediately follows. 
\end{proof}

\begin{theorem}\label{th:chooseinduceditem}
Let $x$ be a parameterized item such that 
\[x = (\nterminal{N}_{f_{v+1}} \rightarrow \gsymbol{X}_1^{f_1} \ldots \gsymbol{X}_d^{f_d} \itemdot \gsymbol{X}_{d+1}^{f_{d+1}} \ldots \gsymbol{X}_v^{f_v}, i, j, \itemparams)\]
(and $v \geq d + 1$). Then for each $\gamma \in \parameters$ there is $x' \in \induced{x}$ with
\[ x' = (\nterminal{N}^{\itemparams_0}_\beta \rightarrow \gsymbol{X}^{\itemparams_1}_{\itemparams_2} \ldots \gsymbol{X}^{\itemparams_{2d-1}}_{\itemparams_{2d}} 
\itemdot \gsymbol{X}_{\gamma}^{\itemparams_{2d+1}} w, i, j)\]
for some $\beta \in \parameters$ and some $w \in (\induced{\nonterminals} \cup \induced{\terminals})^*$.
\end{theorem}
\begin{proof}
The proof is basically the same as the proof of Theorem~\ref{th:nonemptyinduceditems}, only that instead of extending $\itemparams$ we extend $\itemparams\, \gamma$ this time. 
\end{proof}

\begin{theorem}\label{th:findparameterizeditem}
Assume $(\nterminal{N}^{\,\alpha}_\beta \rightarrow w) \in \induced{\rules}$.  
If  $w$ ends with $\failsymbol$ then this implies $|w| \geq 2$.
Furthermore, for any $d$ with
\[0 \leq d \leq \begin{cases} |w| & \text{if $w$ does not end with $\failsymbol$} \\ |w| - 2 &\text{if $w$ ends with $\failsymbol$}\end{cases}\] 
and for any $0 \leq i \leq j \leq |D|$ there is a parameterized item 
\[ x = (\nterminal{N}_{f_{k+1}} \rightarrow \gsymbol{X}^{f_1} \ldots \gsymbol{X}^{f_{k}}, d, i, j, \itemparams) \]
such that $(\nterminal{N}^{\,\alpha}_\beta \rightarrow w, d, i, j) \in \induced{x}$. 
\end{theorem} 
\begin{proof}
Assume $w$ does not end with $\failsymbol$. Then there is a parameterized rule 
$r = \nterminal{N}_{f_{|w|+1}} \rightarrow \gsymbol{X}_1^{f_1} \ldots \gsymbol{X}_{|w|}^{f_{|w|}} \in \rules $
such that $(\nterminal{N}^\alpha_\beta \rightarrow w) \in \induced{r}$, and there are $\alpha_1, \beta_1, \ldots, \alpha_{|w|}, \beta_{|w|}$ such that 
$w = {\gsymbol{X}_1}_{\beta_1}^{\alpha_1} \ldots {\gsymbol{X}_{|w|}}_{\beta_{|w|}}^{\alpha_{|w|}}$ and 
$\alpha\, \alpha_1\,\beta_1 \ldots \alpha_{|w|}\, \beta_{|w|}\, \beta \in \paramhull{f_1, \ldots, f_{|w|+1}}$. Setting 
$x = (r, d, i, j, \itemparams)$ where $\itemparams = \operatorname{take}_{2(d+1)}\,(\alpha\, \alpha_1\,\beta_1 \ldots \alpha_{|w|}\, \beta_{|w|}\, \beta)$ 
 yields the desired result.

On the other hand, assume $w$ ends with $\failsymbol$. Then there is a parameterized rule 
$r = \nterminal{N}_{f_{k+1}} \rightarrow \gsymbol{X}_1^{f_1} \ldots \gsymbol{X}_{k}^{f_{k}} \in \rules$ 
such that $(\nterminal{N}^\alpha_\beta \rightarrow w) \in \induced{r}$ and $1 \leq |w| - 1\leq k$. In particular there are 
$\alpha_1, \beta_1, \ldots, \alpha_{|w|-1}, \beta_{|w|-1}$ such that $w = \gsymbol{X}_{\beta_1}^{\alpha_1} \ldots \gsymbol{X}_{\beta_{|w|-1}}^{\alpha_{|w|-1}}\,\failsymbol$ 
and $\alpha\,\alpha_1\, \beta_1 \ldots \alpha_{|w|-1} \in \paramhull{f_1, \ldots, f_{|w|-1}}$. 
Setting $x = (r, d, i, j, \itemparams)$ where $\itemparams = \operatorname{take}_{2(d+1)}\,(\alpha\, \alpha_1\,\beta_1 \ldots \alpha_{|w|-1})$ 
 yields the desired result in this case as well.
\end{proof}

\begin{theorem}\label{th:sim:Predict}
Assume $\someitems \sim \someitems'$. Then 
\[\operatorname{Predict} k\,\someitems \sim \operatorname{Predict}' k\,\someitems'\]
\end{theorem}
\begin{proof}
We first prove $\induced{\operatorname{Predict} k\,\someitems} \subseteq \norm{\operatorname{Predict}' k\,\someitems'}$, so assume 
$x' \in \induced{\operatorname{Predict} k\,\someitems}$. Then there must be $x \in \operatorname{Predict} k\,\someitems$ such that 
$x' \in \induced{x}$. If $x \in \someitems$ then because of $\someitems \sim \someitems'$ it automatically follows that 
$x' \in \norm{\someitems'} \subseteq \norm{\operatorname{Predict}' k\,\someitems'}$. So assume $x \notin \someitems$. This means that 
\[x = (\nterminal{M}_{g_{u+1}} \rightarrow \itemdot \gsymbol{Y}_1^{g_1} \ldots \gsymbol{Y}_u^{g_u}, k, k, 
 \itemparams_{2d+1}\  g_1(\itemparams_{2d+1}))\] and there is a parameterized item $y$ where
 \[y = (N_{f_{v+1}} \rightarrow \gsymbol{X}_1^{f_1} \ldots \gsymbol{X}_d^{f_d} \itemdot \gsymbol{X}_{d+1}^{f_{d+1}} \ldots \gsymbol{X}_v^{f_v}, 
i, k, \itemparams) \in \someitems\] such that $\gsymbol{X}_{d+1} = \nterminal{M}$. From $x' \in \induced{x}$ it follows that
\[ x' = (\nterminal{M}_{\gamma}^{\itemparams_{2d+1}} \rightarrow \itemdot a, k, k)\]
for some $\gamma \in \parameters$ and some $a \in (\induced{\nonterminals} \cup \induced{\terminals})^*$. From Theorem~\ref{th:chooseinduceditem} we obtain
an item $y'$ with
\[ y' =  (N_\beta^{\itemparams_0} \rightarrow \gsymbol{X}^{\itemparams_0}_{\itemparams_1} \ldots \gsymbol{X}_{\itemparams_{2d}}^{\itemparams_{2d-1}} 
\itemdot \gsymbol{M}_{\gamma}^{\itemparams_{2d+1}}\, w, i, k)\]
for some $\beta \in \parameters$ and some $w \in (\induced{\nonterminals} \cup \induced{\terminals})^*$. Because $\someitems \sim \someitems'$ implies
$y' \in \someitems'$ it follows immediately that $x' \in \operatorname{Predict}' k\, \someitems'$. As $x'$ is an induced item, this implies
$x' \in \norm{\operatorname{Predict}' k\, \someitems'}$.

Now we prove $\norm{\operatorname{Predict}' k\,\someitems'} \subseteq \induced{\operatorname{Predict} k\,\someitems}$. Assume 
$x' \in \norm{\operatorname{Predict}' k\,\someitems'}$. If $x' \in \someitems'$ then $x' \in \induced{\someitems}$ and we are done, so assume $x' \notin \someitems'$.
This means that $x' = (\nterminal{M}^\alpha_\gamma \rightarrow \itemdot a, k, k)$ for some $\nterminal{M}^\alpha_\gamma \rightarrow a \in \induced{\rules}$,
and there is $y' = (\nterminal{N}^\beta_\delta \rightarrow b \itemdot \nterminal{M}^\alpha_\gamma, i, k) \in \someitems'$. This means that there is 
$y \in \someitems$ with $y' \in \induced{y}$. From Theorem~\ref{th:findparameterizeditem} we obtain an $x$ with $x' \in \induced{x}$ and it follows straight
from the definitions that $x \in \operatorname{Predict} k\, \{y\}$.
\end{proof}

\begin{theorem}\label{th:sim:Complete}
Assume $\someitems \sim \someitems'$. Then 
\[\operatorname{Complete} k\,\someitems \sim \operatorname{Complete}' k\,\someitems'\]
\end{theorem}
\begin{proof}
We first prove $\induced{\operatorname{Complete} k\,\someitems} \subseteq \norm{\operatorname{Complete}' k\,\someitems'}$, so assume 
$x' \in \induced{\operatorname{Complete} k\,\someitems}$. Then there must be $x \in \operatorname{Complete} k\,\someitems$ such that 
$x' \in \induced{x}$. We can assume $x \notin \someitems$ as otherwise the result follows immediately. This means that there are parameterized items $y$ and $z$
such that $x$, $y$ and $z$ have the shape
\begin{align*}
 x &=  (N_{f_{v+1}} \rightarrow \gsymbol{X}_1^{f_1} \ldots \gsymbol{X}_d^{f_d} \gsymbol{X}_{d+1}^{f_{d+1}}\itemdot 
 \ldots \gsymbol{X}_v^{f_v}, i, k, 
 \itemparams'\, f_{d+2}(\itemparams')) \\
 y &= (N_{f_{v+1}} \rightarrow \gsymbol{X}_1^{f_1} \ldots \gsymbol{X}_d^{f_d} \itemdot \gsymbol{X}_{d+1}^{f_{d+1}} \ldots \gsymbol{X}_v^{f_v}, i, j, \itemparams) \in \someitems\\
 z &= (\nterminal{M}_{g_{u+1}} \rightarrow \gsymbol{Y}_1^{g_1} \ldots \gsymbol{Y}_u^{g_u} \itemdot, j, k, \altitemparams) \in \someitems
\end{align*}
such that $\gsymbol{X}_{d+1} = M$ and $\itemparams_{2d+1} = \altitemparams_{0}$
and $\itemparams' = \itemparams\, \altitemparams_{2u+1}$.
Moving the dot in $x'$ one position to the left results in an item $y'$, and obviously $y' \in \induced{y}$.
From Theorem~\ref{th:nonemptyinduceditems} we obtain $z' \in \induced{z}$. Because of $\someitems \sim \someitems'$ we have
$\{y', z'\} \subseteq \norm{\someitems'}$. Furthermore it is easy to see that 
$x' \in \operatorname{Complete}' k \, \{y', z'\}$. This concludes this direction of the proof. 

For the other direction  
assume $x' \in \norm{\operatorname{Complete}' k\,\someitems'}$. We need to show that $x' \in \induced{\operatorname{Complete} k\,\someitems}$.
We can assume $x' \notin \someitems'$ as otherwise $x' \in \induced{\someitems}$ follows immediately. Then there exist $y', z' \in \someitems'$
such that $x'$, $y'$ and $z'$ have the shape
\begin{align*}
x' & = (\nterminal{N}^\alpha_\beta \rightarrow \gsymbol{X}^{\alpha_1}_{\beta_1} \ldots \gsymbol{X}^{\alpha_d}_{\beta_d}\,\gsymbol{X}^{\alpha_{d+1}}_{\beta_{d+1}}\itemdot a, i, k)\\
y' & = (\nterminal{N}^\alpha_\beta \rightarrow \gsymbol{X}^{\alpha_1}_{\beta_1} \ldots \gsymbol{X}^{\alpha_d}_{\beta_d} \itemdot \gsymbol{X}^{\alpha_{d+1}}_{\beta_{d+1}}\,a, i, j)\\
z' & = (\gsymbol{X}^{\alpha_{d+1}}_{\beta_{d+1}} \rightarrow b \itemdot, j, k)
\end{align*}
where $a \neq \failsymbol$ as $x' \in \norm{\ldots}$. Because of $\someitems \sim \someitems'$ there exist $y, z \in \someitems$ with $y' \in \induced{y}$, $z' \in \induced{z}$ and
\begin{align*}
 y &= (N_{f_{v+1}} \rightarrow \gsymbol{X}_1^{f_1} \ldots \gsymbol{X}_d^{f_d} \itemdot \gsymbol{X}_{d+1}^{f_{d+1}} \ldots \gsymbol{X}_v^{f_v}, i, j, \itemparams)\\
 z &= (\nterminal{M}_{g_{u+1}} \rightarrow \gsymbol{Y}_1^{g_1} \ldots \gsymbol{Y}_u^{g_u} \itemdot, j, k, \altitemparams) 
\end{align*}
Because of $a \neq \failsymbol$ we know that $f_{d+2}$ is defined on $\itemparams' = \itemparams\, \beta_{d+1}$ and thus $x \in \operatorname{Complete} k\, \{y, z\}$ where
\[ x = (N_{f_{v+1}} \rightarrow \gsymbol{X}_1^{f_1} \ldots \gsymbol{X}_d^{f_d} \gsymbol{X}_{d+1}^{f_{d+1}} \itemdot \ldots \gsymbol{X}_v^{f_v}, i, k, \itemparams'\,f_{d+2}(\itemparams')) \] 
It is clear that $x' \in \induced{x}$, which concludes the proof.
\end{proof}

\begin{theorem}\label{th:sim:Tokens}
Assume $\someitems \sim \someitems'$. Then 
\[\tokenbij(\operatorname{Tokens} T\,k\,\someitems) = \operatorname{Tokens}'\, \tokenbij(T)\,k\,\someitems'\]
\end{theorem}
\begin{proof}
Let us first introduce two abbreviations $V$ and $V'$: 
\begin{align*}
V =\ & \{ x \mid (N_{f_{v+1}} \rightarrow \gsymbol{X}_1^{f_1} \ldots \gsymbol{X}_d^{f_d} \itemdot \gsymbol{X}_{d+1}^{f_{d+1}} \ldots \gsymbol{X}_v^{f_v},
i, k, \itemparams) \in \someitems \\
     & \phantom{\{} \wedge \gsymbol{X}_{d+1} \in \terminals \wedge x \in \lexer(\gsymbol{X}_{d+1}, \itemparams_{2d+1}, D, k)\}\\
V' =\ & \{ x \mid \ X \in \induced{\terminals}\  \wedge (Y \rightarrow a \itemdot X b, i, k) \in \someitems'\\
     &  \phantom{\{} \wedge x \in \induced{\lexer}(X)(D, k)\}
\end{align*}
Unfolding the definitions of $\operatorname{Tokens}$ and $\operatorname{Tokens}'$ shows that the theorem statement is equivalent to
\[ \tokenbij(\selector(T, V)) = \induced{\selector}(\tokenbij(T), V') \]
which according to the definition of $\induced{\selector}$ is equivalent to
\[ \induced{\selector}(\tokenbij(T), \tokenbij(V)) = \induced{\selector}(\tokenbij(T), V'). \]
This means that if we can prove
\[ \tokenbij(V) = V'\]
we have proven the theorem.

We first prove $\tokenbij(V) \subseteq V'$, so assume $(t^\alpha_\beta, c) \in \tokenbij(V)$, or equivalently
$(t, \alpha, \beta, c) \in V$. This means that there is a parameterized item 
\[x = (N_{f_{v+1}} \rightarrow \gsymbol{X}_1^{f_1} \ldots \gsymbol{X}_d^{f_d} \itemdot \gsymbol{X}_{d+1}^{f_{d+1}} \ldots \gsymbol{X}_v^{f_v},
i, k, \itemparams) \in \someitems \]
with $\gsymbol{X}_{d+1} = t$, $\itemparams_{2d+1} = \alpha$ and $(t, \alpha, \beta, c) \in \lexer(t, \alpha, D, k)$. 
Theorem~\ref{th:chooseinduceditem} yields an item $x' \in \induced{x} \subseteq \induced{\someitems} \subseteq \someitems'$ such that
\[ x' = (\nterminal{N}^{\itemparams_0}_\delta \rightarrow \gsymbol{X}^{\itemparams_1}_{\itemparams_2} \ldots \gsymbol{X}^{\itemparams_{2d-1}}_{\itemparams_{2d}} 
\itemdot t_{\beta}^{\alpha}\, w, i, k)\]
for some $\delta$ and some $w$. This proves $(t^\alpha_\beta, c) \in V'$. 

To prove $V' \subseteq \tokenbij(V)$, assume $(t^\alpha_\beta, c) \in V'$. Then there must be an item 
\[x' = (Y \rightarrow a \itemdot t^\alpha_\beta\, b , i, k) \in \someitems' \]
such that $(t^\alpha_\beta, c) \in \induced{\lexer}(t^\alpha_\beta)(D, k)$, or equivalently 
\[(t, \alpha, \beta, c) \in \lexer(t, \alpha, D, k)\]
Because the dot in $x'$ appears before a terminal symbol we know that $x' \in \norm{\someitems'}$, thus there is an $x \in \someitems$ with
\[x = (N_{f_{v+1}} \rightarrow \gsymbol{X}_1^{f_1} \ldots \gsymbol{X}_{|a|}^{f_{|a|}} \itemdot \gsymbol{X}_{|a|+1}^{f_{|a|+1}} \ldots \gsymbol{X}_v^{f_v}, i, k, \itemparams)\]
such that $x' \in \induced{x}$. This implies $\gsymbol{X}_{|a|+1} = t$ and $\itemparams_{2|a|+1} = \alpha$. This proves $(t, \alpha, \beta, c) \in V$. 
\end{proof}

\begin{theorem}\label{th:sim:Scan}
Assume $\someitems \sim \someitems'$. Then 
\[\operatorname{Scan} T\, k\,\someitems \sim \operatorname{Scan}' \tokenbij(T)\, k\,\someitems'\]
\end{theorem}
\begin{proof}
The proof is straightforward and similar to the proof of Theorem~\ref{th:sim:Complete}.
\end{proof}

\newcommand{\itemsiprime}[1]{\ensuremath{{\mathcal{I}}'_{#1}}}
\newcommand{\itemsjprime}[2]{\ensuremath{{\mathcal{J}'}_{#1}^{\,#2}}}
\newcommand{\ctokensprime}[2]{\ensuremath{{\mathcal{T}'}_{#1}^{\,#2}}}
\newcommand{\allitemsprime}{\ensuremath{\mathfrak{I}'}}

\begin{figure}
\begin{align*}
\pi'_k\,T\,\someitems & = \limit {(\operatorname{Scan}' T\ k \circ \operatorname{Complete}' k \circ \operatorname{Predict}' k)}{\someitems}\\
\itemsjprime 0 0 & = \pi'_0\ \emptyset\ \operatorname{Init}'\\
\itemsjprime k {u+1} & = \pi'_k\ \ctokensprime k {u+1}\ \itemsjprime k u\\
\itemsiprime k & = \bigcup\limits_{u=0}^\infty \itemsjprime k u\\
\itemsjprime {k+1} 0 & = \pi'_{k+1}\ \emptyset{}\ \itemsiprime k\\
\ctokensprime k 0 & = \emptyset\\
\ctokensprime {k} {u+1} & = \operatorname{Tokens}'\ \ctokensprime k u \  k\ \itemsjprime k u\\
\allitemsprime & = \itemsiprime {|D|}
\end{align*}
\caption{ELLA equations}
\label{fig:ella}
\end{figure}

Apart from using parameterized building blocks, PELLA as shown in Figure~\ref{fig:pella} is identical to ELLA as shown in Figure~\ref{fig:ella}. Given that we now know that
all of PELLA's building blocks simulate those of ELLA as described in Figure~\ref{fig:simulation}, it is easy to see that the relationship $\allitems \sim \allitems'$ holds for the 
result $\allitems$ of PELLA and the result $\allitemsprime$ of ELLA. We show this in the following through a sequence of simple theorems.

\begin{theorem}\label{th:sim:bigcup}
Assume $\someitems_u \sim \someitems'_u$ for all $u \in \{0, 1, 2, \ldots \}$. Then 
\[ \left(\bigcup\limits_{u=0}^\infty \someitems_u\right) \sim \left(\bigcup\limits_{u=0}^\infty \someitems'_u\right) \]
\end{theorem}
\begin{proof}
\[
\induced{\bigcup\limits_{u=0}^\infty \someitems_u} = \bigcup\limits_{u=0}^\infty \induced{\someitems_u} = 
\bigcup\limits_{u=0}^\infty \norm{\someitems'_u} = \norm{\bigcup\limits_{u=0}^\infty \someitems'_u}
\]
\end{proof}

We say that a function $f$ simulates a function $g$ iff $\someitems \sim \someitems'$ implies $f\,\someitems \sim g\,\someitems'$ for any $\someitems$ and $\someitems'$.

\begin{theorem}\label{th:sim:composition}
If $f$ simulates $f'$ and $g$ simulates $g'$ then $g \circ f$ simulates $g' \circ f'$. 
\end{theorem}
\begin{proof}From $\someitems \sim \someitems'$ follows $f\,\someitems \sim f'\,\someitems'$, and from that
$g\,(f\,\someitems) \sim g'\,(f'\,\someitems')$ follows. \end{proof}

\begin{theorem}\label{th:sim:power}
Assume that $f$ simulates $g$. Then for any $n \in \{0, 1, 2, \ldots\}$ we have that $f^n$ simulates $g^n$.
\end{theorem}
\begin{proof}This is an immediate consequence of Theorem~\ref{th:sim:composition}.\end{proof}

\begin{theorem}\label{th:sim:limit}
Assume that $f$ simulates $g$. Then $\operatorname{limit}\, f$ simulates $\operatorname{limit}\,g$.
\end{theorem}
\begin{proof}
We need to show that from $\someitems \sim \someitems'$ it follows that
\[  \left(\bigcup\limits_{n=0}^\infty f^n(\someitems)\right) \sim \left(\bigcup\limits_{n=0}^\infty g^n(\someitems')\right)     \]
which follows immediately from Theorems~\ref{th:sim:bigcup} and~\ref{th:sim:power}.
\end{proof}

\begin{theorem}\label{th:sim:pi}
$\pi_k\, T$ simulates $\pi'_k\ \tokenbij(T)$.
\end{theorem}
\begin{proof}
We know that $\operatorname{Predict}\,k$ simulates $\operatorname{Predict}'\,k$ (Theorem~\ref{th:sim:Predict}), 
$\operatorname{Complete}\,k$ simulates $\operatorname{Complete}'\,k$ (Theorem~\ref{th:sim:Complete}) and
$\operatorname{Scan}\,T\,k$ simulates $\operatorname{Scan}'\,\tokenbij(T)\, k$ (Theorem~\ref{th:sim:Scan}).
From this and Theorems~\ref{th:sim:composition} and ~\ref{th:sim:limit} the result follows immediately.
\end{proof}

\begin{theorem}\label{th:sim:base}
$\itemsj 0 0 \sim \itemsjprime 0 0$
\end{theorem}
\begin{proof}
We have $\pi'_0\,\emptyset\, \operatorname{Init}' = \pi'_0\,\emptyset\,(\operatorname{Predict}' 0\,\operatorname{Init}')$, which together
with Theorems~\ref{th:sim:Init} and~\ref{th:sim:pi} implies $\itemsj 0 0 \sim \itemsjprime 0 0$.
\end{proof}

\begin{theorem}\label{th:simulate:helper}
Assume $\itemsj k 0 \sim \itemsjprime k 0$. Then
\begin{align*}
\itemsj k u &\sim \itemsjprime k u \quad\text{for all $u \in \{0, 1, 2, \ldots\}$} \\
\tokenbij(\ctokens k u) &= \ctokensprime k u \quad\text{for all $u \in \{0, 1, 2, \ldots\}$}\\
\itemsi k &\sim \itemsiprime k 
\end{align*}
\end{theorem}
\begin{proof}
We first prove $\itemsj k u \sim \itemsjprime k u$ and $\tokenbij(\ctokens k u) = \ctokensprime k u$ by induction over $u$. The base case $u=0$ follows trivially
from our assumption $\itemsj k 0 \sim \itemsjprime k 0$ and the fact that $\tokenbij(\emptyset) = \emptyset$. For the induction step $u \rightarrow u + 1$
we apply Theorem~\ref{th:sim:Tokens} and the induction hypothesis and obtain $\tokenbij(\ctokens k {u + 1}) = \ctokensprime k {u + 1}$. From this, together with 
Theorem~\ref{th:sim:pi} and the induction hypothesis, 
we obtain $\itemsj k {u+1} \sim \itemsjprime k {u+1}.$ 

From $\itemsj k u \sim \itemsjprime k u \quad\text{for all $u \in \{0, 1, 2, \ldots\}$}$ and Theorem~\ref{th:sim:bigcup} follows then immediately $\itemsi k \sim \itemsiprime k$.
\end{proof}

\begin{theorem}\label{th:simulate:main}
\begin{align*}
\itemsj k u &\sim \itemsjprime k u \quad\text{for all $k, u \in \{0, 1, 2, \ldots\}$} \\
\tokenbij(\ctokens k u) &= \ctokensprime k u \quad\text{for all $k, u \in \{0, 1, 2, \ldots\}$}\\
\itemsi k &\sim \itemsiprime k \quad\text{for all $k$}\\
\allitems &\sim \allitemsprime
\end{align*}
\end{theorem}
\begin{proof}
The first three formulas follow immediately by induction over $k$ and application of Theorem~\ref{th:sim:base} and Theorem~\ref{th:simulate:helper} for the base case, and Theorem~\ref{th:sim:pi}
and Theorem~\ref{th:simulate:helper} for the induction step, respectively.

We then deduce $\allitems = \itemsi {|D|} \sim \itemsiprime {|D|} = \allitemsprime$.
\end{proof}

\newcommand{\generateditems}[1]{\ensuremath{\langle{#1}\rangle}}

We are now in a position to prove the correctness of PELLA.
\begin{proof}[Proof of Theorem~\ref{th:correctness:PELLA}]
We first introduce the abbreviation 
\begin{align*}
P = \{ \itemparams_{2 u + 1} \mid & (\startsymbol_{f_{u+1}} \rightarrow \gsymbol{X}_1^{f_1} \ldots \gsymbol{X}_u^{f_u} \itemdot, 0, |D|, \itemparams) \in \allitems \\
 & \wedge \itemparams_0 = \startparameter \}
\end{align*}
Then our goal is to prove $\outputs{D} = P$. 

To prove $\outputs{D} \subseteq P$, assume $\beta \in \outputs{D}$. Then there is a path $p \in \locallexing(D)$ such that 
$\startnonterminal \Rightarrow \startsymbol^{\startparameter}_\beta \derives \terminalsof{p}$. This means that there is an item
$x' = (\startsymbol^{\startparameter}_\beta \rightarrow w, 0, |D|)$ which is $p$-valid, i.e. $x' \in \generateditems{\allpaths}$.
From the correctness proof of ELLA we know that $\generateditems{\allpaths} = \allitemsprime$ and thus $x' \in \allitemsprime$, and the shape of $x'$ implies even $x' \in \norm{\allitemsprime}$. Therefore Theorem~\ref{th:simulate:main} allows us to deduce $x' \in \induced{\allitems}$, and this
means that there is an $x \in \allitems$ such that $x' \in \induced{x}$ and such that $x$ has the shape
\[ x =  (\startsymbol_{f_{u+1}} \rightarrow \gsymbol{X}_1^{f_1} \ldots \gsymbol{X}_u^{f_u} \itemdot, 0, |D|, \itemparams)  \]
From $x' \in \induced{x}$ follows immediately that $\itemparams_0 = \startparameter$ and $\itemparams_{2u+1} = \beta$, and therefore
$\beta \in P$.

To prove the other direction, $P \subseteq \outputs{D}$, assume we have an $x \in \allitems$ such that
$x =  (\startsymbol_{f_{u+1}} \rightarrow \gsymbol{X}_1^{f_1} \ldots \gsymbol{X}_u^{f_u} \itemdot, 0, |D|, \itemparams)$
and $\itemparams_0 = \startparameter$. We need to prove $\itemparams_{2u+1} \in \outputs{D}$. Theorem~\ref{th:nonemptyinduceditems} tells us that
there is an $x' \in \induced{x}$, and this $x'$ has necessarily the form 
\[ x' = (\startsymbol^{\startparameter}_{\itemparams_{2u+1}} \rightarrow w \itemdot, 0, |D|)\]
From Theorem~\ref{th:simulate:main} and the correctness of ELLA follows 
$x' \in \induced{x} \subseteq \induced{\allitems} \subseteq \allitemsprime = \generateditems{\allpaths}$. 
Thus there is a path $p \in \allpaths$ such that $x'$ is $p$-valid, i.e. there is $v \in \{0, \ldots, |p| \}$ such that
\begin{align*}
  |\charsof{p}| &= |D| \\
  |\charsof{p_0 \ldots p_{v-1}}| &= 0\\
  \startnonterminal & \derives \terminalsof{p_0 \ldots p_{v-1}} \startsymbol^{\startparameter}_{\itemparams_{2u+1}} \gamma 
    \quad\text{for some $\gamma$}\\
  w &\derives \terminalsof{p_v \ldots p_{|p|-1}}
\end{align*} 
As part of the correctness proof of ELLA we have previously seen that if we drop the first $v$ empty tokens of $p$, the result
$q = p_v \ldots p_{|p|-1}$ will still be in $\allpaths$ (Theorem 5.2 in~\cite{locallexing}). We deduce 
\[ \startnonterminal \Rightarrow \startsymbol^{\startparameter}_{\itemparams_{2u+1}} \Rightarrow w \derives \terminalsof{q}\]
and $q \in \locallexing(D)$, and thus $\itemparams_{2u+1} \in \outputs{D}$.
\end{proof}

%\section{Parse Trees}

%\section{Practical Implementation}

\bibliographystyle{abbrvnat}

\end{document}